\documentclass{article}[12pt] 
\usepackage{braket}
\usepackage{amsmath}
\usepackage{amsfonts}
\usepackage[mathscr]{eucal} 
\usepackage{amsmath,amssymb,graphicx}  
\usepackage{graphicx}  
\usepackage{bm}       
\usepackage{color}
\usepackage{mathtools} 
\usepackage{sidecap}
\usepackage{wrapfig}    
\usepackage{subfig} 
\usepackage{float} 
\begin{document} 
\title{A Magnus approximation approach to harmonic systems with 
time-dependent frequencies}
\author{C.~D.~Fosco$^{a}$, F.~C.~Lombardo$^{b}$ and F.~D.~ Mazzitelli$^{a}$ \\ \\
$^a$Instituto Balseiro and Centro At\'omico Bariloche, 
R8402AGP Bariloche, Argentina\\ \\
 $^b$ Departamento de F\'\i sica {\it Juan Jos\'e Giambiagi} ,  FCEyN UBA and IFIBA  \\
CONICET-UBA, Facultad de Ciencias Exactas y Naturales, \\
 Ciudad Universitaria, Pabell\' on I, 1428 Buenos Aires, Argentina }

%

\maketitle

\begin{abstract} 
\noindent 
We use a Magnus approximation at the level of the equations of motion for a harmonic
system with a time-dependent frequency,  to find an expansion for its  {\em
in-out\/} effective action,  and a unitary expansion for the Bogoliubov
transformation between {\em in\/} and {\em out\/} states. 
The dissipative effects derived therefrom are compared with the ones obtained from perturbation
theory in powers of the time-dependent piece in the frequency, and with those
derived using multiple scale analysis in systems with parametric resonance. 
We also apply the Magnus expansion to the {\em in-in\/} effective action, to
construct reality and causal equations of motion for the external system. We
show that the nonlocal equations of motion can be written in terms of a ``retarded
Fourier transform" evaluated at the resonant frequency.
\end{abstract}
\maketitle
\section{Introduction}\label{sec:intro}
Harmonic oscillators with time-dependent frequencies are ubiquitous in many
branches of physics. In the context of quantum field theory, there are many
examples where the modes of a free field, when put under the influence of
time-dependent external conditions, can be described as a system of harmonic
oscillators with time-dependent frequencies and couplings.  A well-known
example arises when considering quantum fields in cosmological backgrounds,
such that partial homogeneity implies uncoupled field modes which are harmonic
oscillators, with time-dependent frequencies~\cite{Birrell}. Another celebrated
example, which produces {\em coupled\/} modes, corresponds to quantum fields in
the presence of moving mirrors, or time-dependent media, which can also be
treated as a set of coupled harmonic oscillators (see, for
example  \cite{Reviews,nos1,nos2} and references therein). 

The most interesting feature of the system is, perhaps, that it exhibits, at
the quantum level, a parametric resonance phenomenon. This can be studied, for
example, in terms of the Bogoliubov transformation between the {\it in} and
{\it out} states, or the {\it in-out} effective action, both of which make it
possible to obtain the particle creation rates. In terms of a given mode, those
rates correspond to the creation of a certain number of quanta, as in the
dynamical Casimir effect \cite{Reviews}.  The same physical phenomenon is also manifested when
considering another observable, namely, the dynamical equations for the
external degrees of freedom. They may be obtained from the {\it in-in}
effective action, and exhibit a back-reaction due to emission of quantum field
modes~\cite{CalHu94,nos3}.  

Except for rather special cases, it is not possible to obtain closed
expressions for the effective action, even in the case of a single harmonic
mode with a time-dependent frequency. Indeed, the problem of evaluating
the effective action for a system like this, may be posed in terms of a
functional determinant, an object which may be obtained from the solution
of a linear second-order differential equation:
the classical equation of motion.  Since, except for rather special cases,
closed solutions for the latter are not known, it is natural to implement
approximate treatments. Assuming that the time dependent piece of the
frequency is small in comparison with the constant (average) part, an
expansion in powers of the former seems natural. The alternative we follow
here, corresponds to writing the  (formal) solution to the classical
equation of motion in phase space, and implementing the Magnus expansion
\cite{Magnus,11,12} to solve the latter. This preserves, order by order, the time evolution as a
canonical transformation, both at the classical and quantum
levels (since the equations of motion are linear). This is a higher desirable 
feature when considering an expansion for the Bogoliubov transformations,
Thus, our approach may be interpreted as an alternative expansion in the 
time-dependent part of the frequency, which preserves unitarity, and would
correspond to a resummation of infinite terms on the usual expansion.

The Magnus approach has been used in a large number of works and is of interest 
in many branches of quantum mechanics. For example, it han been used in the 
open quantum systems theory as a tool to investigate how to manipulate 
the irreversible component of open-system evolutions (decoherence and 
dissipation) through the 
application of external  controllable interactions \cite{viola}. Also in the 
study of periodically driven systems, by means of an expansion both in the driving 
term and the inverse of the driving frequency, applicable to isolated or dissipative 
systems. In \cite{parametric}, a systematic Magnus expansion is used to derive explicit expressions 
for a system with a driving term with harmonic time dependence. Also, in Ref. \cite{bloch} authors 
evaluate the squeezing parameters and the corresponding squeezing eigenmodes in the 
frame of the Bloch-Messiah decomposition for the broadband squeezed light generated 
by type-I parametric down-conversion with monochromatic pump. Magnus expansion provides 
the first three approximation orders for  the quantum evolution operator. 

In Ref. \cite{sipe} it is shown that even there is a wide range of problems where an elementary 
first order perturbative approach would fail, but a calculation using just the first order 
Magnus term would be sufficient. They describe both pump pulse and phase-matching functions that enter 
into the description of the photon generation or conversion (such as spontaneous parametric down-conversion and 
wave mixing). 

Recently, the dynamics of a chirped two-level system has been studied. Authors in \cite{2LS} derived 
a Magnus expansion for the Hamiltonian which determines the stroboscopic \cite{thimmel} dynamics of a non-harmonically driven 
two-level system, i.e. for a linear frequency chirp.   

The paper represent a pedagogical presentation of the effective action approach for 
harmonic systems with time-dependent frequency, and it is organized as follows: in Sect.~\ref{sec:osc} we write the
{\it in-out} effective action for a single harmonic oscillator with a
time-dependent frequency, in terms of an evolution operator. The approach is based on the
Gelfand-Yaglom's (GY) theorem~\cite{Gelfand} to evaluate functional
determinants. In Sect. \ref{sec:Magnus} we apply the above mentioned Magnus
expansion to the solution of the homogeneous second-order equation which
yields the result for the fluctuation determinant.  We evaluate explicitly
the first and second order terms in that expansion, and present the
structure of the third order one.  By studying the imaginary part of the
corresponding terms in the effective action, we show that it automatically
captures non-trivial features, like the position of the (parametric)
resonances.  We also show that, in this approximation, the leading order
in the Magnus expansion involves the vacuum expectation value of a squeeze
operator. In Sect. ~\ref{sec:ctp} we consider the {\it in-in} effective action,
implementing the Magnus approximation also, at the level of the
equations of motion. We study the implementation of the properties of
reality and causality within the context of this approximation.
Sect.~\ref{sec:conc} contains our conclusions. 
\section{Harmonic oscillator with a time-dependent
frequency}\label{sec:osc}
\subsection{The system and its effective action}
Let us begin by considering a system which exhibits the simplest
non-trivial realization of the phenomenon of parametric resonance; namely,
that of a single quantum degree of freedom, endowed with the dynamics of a
harmonic oscillator, with a time-dependent frequency $\omega(t)$.  Denoting
by $q$ its associated generalized coordinate, its
classical action ${\mathcal  S}$ is given by:
\begin{equation}
{\mathcal  S}(q) \;=\; \frac{1}{2} \, \int dt \, 
\left[ \dot{q}^2(t)  \,-\, \omega^2(t) q^2(t)  \right] \;.
\end{equation}
We split $\omega^2(t) > 0$ into a positive constant component
$\omega_0^2$, plus a time-dependent part $\epsilon(t)$: 
\begin{equation}
	\omega^2(t) \;=\; \omega_0^2 \,+\, \epsilon(t) \;,
\end{equation}
where $|\epsilon(t)| < \omega_0^2$. The splitting becomes unambiguous when
we impose on $\epsilon(t)$ the condition \mbox{$\int_{-\infty}^{+\infty} dt
\, \epsilon(t) = 0$}. 
The sign of $\omega_0$ is chosen, by convention, to be positive.  

We then introduce the in-out effective action $\Gamma$ (a functional of
$\epsilon(t)$), as the quotient between two path integrals, namely:
\begin{equation}\label{eq:defgamma}
	e^{i \Gamma} \;=\; \frac{\int \, {\mathcal  D}q \; e^{i {\mathcal
	S}(q)}}{\int {\mathcal  D}q \; e^{i {\mathcal  S}_0(q)}} \;,	
\end{equation}
where the denominator, introduced for normalization purposes, has been
defined in terms of the constant-frequency action ${\mathcal  S}_0$:
\begin{equation}
{\mathcal  S}_0(q) \;=\; \frac{1}{2} \, \int dt \, \left[ \dot{q}^2(t)  
\,-\, \omega_0^2 q^2(t)  \right] \;.
\end{equation}

In Eq.(\ref{eq:defgamma}), the functionally integrated paths $q(t)$ are the
appropriate ones in order to calculate the in-out effective action. 
These paths must vanish at infinity; we shall reach that limit by starting
from a finite interval  $t \in [-\frac{T}{2}\,,\, \frac{T}{2}]$, imposing
the boundary conditions: $ q(-T/2) \,=\, q(+T/2) \,=\,0$, and then taking
the $T \, \to \, ( 1 - i 0^+) \, \infty$ limit. 
Note that, because of the imaginary part in the previous limit, no
non-trivial classical solution survives the limiting procedure.

Therefore, and since the functional integrals in (\ref{eq:defgamma}) are
Gaussian, we may express $\Gamma$ just in terms of functional determinants:
\begin{equation}
	e^{i \Gamma} \;=\; 
	\left(\frac{\det {\mathcal  K}}{\det {\mathcal K}_0} \right)^{-1/2} \;,
\end{equation}
where we have introduced the ${\mathcal  K}$ and ${\mathcal  K}_0$ operators
corresponding, respectively, to ${\mathcal S}$ and ${\mathcal S}_0$:
\begin{equation}
{\mathcal  K}  \, \equiv \, \frac{d^2}{dt^2} + \omega^2(t) \;\;,\;\;\;\; 
{\mathcal  K}_0 \, \equiv \frac{d^2}{dt^2} + \omega_0^2  \;.
\end{equation}

Our next step is then to evaluate the ratio between the two functional
determinants which, by a rather straightforward application of
the GY theorem \cite{Gelfand}, may be written as follow, may be written as
follows:
\begin{equation}\label{eq:gy}
\frac{\det {\mathcal  K}}{\det {\mathcal  K}_0} \;=\; \frac{q(T/2)}{q_0(T/2)} \;,
\end{equation}
where $q$ and $q_0$ are the unique solutions to the homogeneous equations:
\begin{equation}
{\mathcal  K}q(t) \,=\,0 \;\;,\;\;\;\; {\mathcal  K}_0q_0(t) \,=\,0 \;\;, 
\end{equation}
for the initial conditions $q(-T/2) = 0$, $p(-T/2)=1$ ($p \equiv \dot{q}$)
(and identical conditions for $q_0$).


\subsection{First-order treatment of the G-Y approach}
The problem has, therefore, been reduced to the calculation of the solution
to a homogeneous second-order equation with prescribed initial conditions \cite{Gelfand}.
It is well-known that a second order equation may be equivalently
formulated as a system of two first-order equations: that reformulated
problem will be the subject of our approximation scheme. To that order, we
first introduce the two complex combinations: 
\begin{equation}
a \;=\; \frac{\omega_0 q + i p}{\sqrt{2 \omega_0}} \;,\;\;\;
a^* \;=\; \frac{\omega_0 q - i p}{\sqrt{2 \omega_0}} 
\end{equation}
and the two-component function
\begin{equation}
| \Psi(t) \rangle \;\equiv\; 
\left( 
\begin{array}{c}
a(t) \\
a^*(t)
\end{array}
\right) \;,
\end{equation}
with the initial condition:
\begin{equation}
|\Psi(-T/2) \rangle = 
\frac{i}{\sqrt{2 \omega_0}} \, \left( 
\begin{array}{c}
1 \\
-1
\end{array}
\right) \;.
\end{equation}
The second-order homogeneous equation for $q$, becomes equivalent to a
Schr\"odinger-like equation, with a non-Hermitian Hamiltonian:
\begin{equation}\label{eq:first}
i \, \frac{d}{dt} |\Psi(t)\rangle \;=\; {\mathcal  H}(t) |\Psi(t)\rangle \;\;,
\;\;\;
{\mathcal  H}(t) \;=\; {\mathcal  H}_0 \,+\, {\mathcal  H}'(t) \;,
\end{equation}
where:
\begin{equation}
{\mathcal  H}_0 \,\equiv\,
\omega_0 \, \left( 
\begin{array}{cc}
1 & 0 \\
0  & -1
\end{array}
\right)
\;\;,
\;\;\;\;
{\mathcal  H}'(t) \,\equiv\, \eta(t) \,
\left( 
\begin{array}{cc}
1 & 1 \\
- 1 & -1
\end{array}
\right) \;\;,
\end{equation}
with $\eta(t) \equiv \frac{\epsilon(t)}{2 \omega_0}$, which has the same
dimensions as $\omega_0$.

Corresponding to some initial time $t_i$,  it is natural to introduce
an `evolution operator' (i.e., in mathematical terminology, 
a `fundamental matrix solution') to the first order equation:
\begin{equation}\label{eq:first1}
i \partial_t {\mathcal  U} (t,t_i)\,=\, 
{\mathcal  H}(t) \, {\mathcal U}(t,t_i) \;,
\;\;\;
{\mathcal  U}(t_i,t_i)\;=\; {\mathcal  I} \;,
\end{equation}
(${\mathcal  I} \equiv$ $2 \times 2$ identity matrix).

We note that the evolution operator may be regarded as the (linear) mapping
between initial condition in phase space, and the value of the dynamical
variables at an arbitrary time.  
An important property emerges as a consequence of the tracelessness of
${\mathcal  H}(t)$, namely, the evolution operator
belongs to the $SL(2,{\mathcal  R})$ group. Indeed, $\det [{\mathbb
U}(t_i,t_i)] = 1$, and:
\begin{align}
\partial_t  \det [{\mathcal  U}(t,t_i)] &=\,   \det [{\mathcal  U}(t,t_i)] \;
{\rm Tr} [{\mathcal  H}(t)] \;= \;0  \nonumber\\
& \Rightarrow \;  \det [{\mathcal  U}(t,t_i)] \;=\;1  \;\;, \;\;\; \forall t
\;. 
\end{align}
This property is a manifestation, in the first-order framework, of the
constancy of the Wronskian of two solutions of the original second-order
problem. In the context of the solution of the classical equations of
motion, this condition amounts to the constancy of the Poisson brackets 
involving $a$, $a^*$. 

Equipped with the previously introduced objects, and defining
\mbox{$| i \rangle \equiv \left(\begin{array}{c} 1 \\ -1 \end{array}
\right)$}, 
$| f \rangle \equiv \left(\begin{array}{c} 1 \\ 1 \end{array} \right)$,  
we see that (\ref{eq:gy}) can be written as follows:
\begin{equation}\label{eq:gyf}
\frac{\det {\mathcal  K}}{\det {\mathcal  K}_0} \;=\;
\frac{{\mathcal U}_{fi}}{{\mathcal U}_{fi}^{(0)}}
\end{equation}
where:
\begin{equation}
	{\mathcal U}_{fi} \,\equiv \, \lim_{T \to \infty (1 - i 0^+)}  \langle f | 
	{\mathcal U}(T/2,-T/2)|i \rangle \;\;, \;\;\;
{\mathcal U}_{fi}^{(0)}  \,\equiv \, {\mathcal U}_{fi}\big|_{\eta \to 0} \;.
\end{equation}

The problem thus reduces to finding either exact or approximate 
expressions for ${\mathcal U}_{fi}$. To derive approximate expressions, we
shall introduce an expansion around the `free', $\epsilon = 0$ solution;
i.e., we treat the constant-frequency part of the evolution exactly. 
This is implemented by using, following the quantum mechanical terminology, an 
interaction representation. Introducing interaction representation vectors
in the standard way: $|\Psi_I(t) \rangle \, \equiv \, [{\mathcal
U}^{(0)}(T/2)]^{-1} \; |\Psi(t) \rangle$, where:
\begin{equation}
	{\mathcal U}^{(0)}(t) \,\equiv\, e^{- i t {\mathcal H}_0 }  
	\,=\, \left( 
\begin{array}{cc}
	e^{- i \omega_0 t}  & 0 \\
	0 & e^{i \omega_0 t} 
\end{array}
\right) \;,
\end{equation}
one obtains:
\begin{equation}
{\mathcal  U}(T/2,-T/2) \;=\; {\mathcal  U}^{(0)}(T/2) \; 
{\mathcal U}_I(T/2,-T/2) \; [{\mathcal  U}^{(0)}(-T/2)]^{-1} \;,
\end{equation}
\begin{equation}
i \partial_t {\mathcal  U}_I (t,t_i)\,=\, 
{\mathcal  H}'_I(t) \, {\mathcal U}_I(t,t_i) \;,
\;\;\;
\end{equation}
where 
\begin{equation}
	{\mathcal  H}'_I(t) \,\equiv\, [{\mathcal  U}^{(0)}(t)]^{-1} \, 
	{\mathcal  H}'(t) \, {\mathcal  U}^{(0)}(t) \;.
\end{equation}

The explicit form of ${\mathcal H}'_I(t)$ for the system at hand is:
\begin{equation}\label{eq:hpi}
{\mathcal  H}'_I(t) \;=\; \eta(t) \, \left( 
\begin{array}{cc}
1 & e^{2 i \omega_0 t} \\
- e^{-2 i \omega_0 t} &  -1 
\end{array}
\right) \;.
\end{equation}

It is worth noting that, by taking the appropriate infinite time limit, we
may give a more explicit form for the ratio between determinants, now in
terms of the interaction representation evolution operator. Indeed,
\begin{equation}\label{eq:gyf2}
	\frac{\det {\mathcal  K}}{\det {\mathcal  K}_0} \;=\; \big[{\mathcal
	U}_I \big]_{22}
\end{equation}
where the subindices denote the respective matrix element in the $2 \times
2$ matrix, and ${\mathcal U}_I \equiv {\mathcal U}_I(+\infty,-\infty)$. 


\section{Magnus expansion}\label{sec:Magnus}
A known approach to the determination of ${\mathcal  U}_I$ that
preserves its $SL(2,{\mathcal  R})$ structure is the Magnus expansion. 
Indeed, writing ${\mathcal  U}_I = e^A$, it yields an expansion for $A$ \cite{pedagogico}. 
Denoting in what follows by $A_i$ the term of order $i$ in the Magnus
expansion for $A$:
\begin{equation}
A \;=\; A_1 \,+\, A_2 \,+ \ldots 
\end{equation}

A very important property of this expansion is that it preserves the
unimodularity of the evolution operator, to each order. As we shall see,
this can be interpreted within the context of Bogoliubov
transformation. 

Let us first consider the results for the effective action which are obtained
to each order in the Magnus expansion.
\subsection{First order}
The first-order term in the Magnus expansion, is:
\begin{equation}
	A_1 \;=\; -i \int_{-\infty}^{+\infty} dt \,{\mathcal  H}'_I(t) \;.
\end{equation} 
Using the explicit form for ${\mathcal  H}'_I(t)$ in (\ref{eq:hpi}), we see
that:
\begin{equation}
	A_1 \;=\; \, \left( 
	\begin{array}{cc}
	0  & -i \tilde{\eta}(- 2 \omega_0)  \\
	i \tilde{\eta}(2 \omega_0) & 0  
\end{array}
\right) \;
\end{equation} 
where the tildes denote Fourier transformation. Note that we have assumed that
$\tilde\eta(0)=0$.
To this order we then have:
\begin{equation}
{\mathcal  U}_I\,=\, e^{A_1} \,=\;
{\mathcal  U}_I^{(1)} \;=\; \cosh|\tilde{\eta}(2\omega_0)| \, {\mathcal  I} \,+\,
\frac{\sinh|\tilde{\eta}(2\omega_0)|}{|\tilde{\eta}(2\omega_0)|} \, A_1 \;.
\end{equation}
Therefore, at the first order,
\begin{equation}
	\big[ {\mathcal  U}_I \big]_{22} \,=\, \cosh|\tilde{\eta}(2\omega_0)|\;.
\end{equation}

Inserting this result into the expression for $\Gamma$, we see that:
\begin{equation}\label{resImGamma}
{\rm Im } \Gamma \;=\; \frac{1}{2} \, \log
\cosh\big[\frac{\tilde{\epsilon}(2 \omega_0)}{2 \omega_0}\big] \;.
\end{equation}
This, whenever the excitation $\epsilon(t)$ has a non-vanishing Fourier
component along $2 \omega_0$, there will be a non-zero probability of
vacuum decay. A special case arises when $\epsilon(t)$ is periodic, with a
frequency equal to, say $\Omega$:
\begin{equation}
\epsilon(t) \;=\; \epsilon_0 \, \cos(\Omega t) \;.
\end{equation}
In this case, we see that there will be no imaginary part unless $\Omega =
2 \omega_0$, and that when that happens,
\begin{equation}
\tilde{\epsilon}(2 \omega_0)=\epsilon_0 T/2 \;,
\end{equation}
and we therefore obtain a non-vanishing vacuum decay probability 
\begin{equation}
\vert\langle 0_{in}\vert 0_{out}\rangle\vert^2=e^{-2 {\rm Im(\Gamma)}}=e^{-\frac{\epsilon_0 T}{4\omega_0}}\, .
\end{equation}
The exponential behavior of this probability is produced by parametric resonance. Similar results can be obtained
using the method of multiple scales \cite{Bender}. Our result is more general, since can be applied
to a generic function $\epsilon(t)$, not only to the particular case of a harmonic perturbation.

\subsection{Second and third orders}
The second order term is given by:
\begin{equation}
A_2 \;=\; - \frac{1}{2} \, \int_{-\infty}^{+ \infty} dt_1
\int_{-\infty}^{t_1} dt_2 \; 
\big[ 
{\mathcal  H}'_I(t_1) \;,\; {\mathcal  H}'_I(t_2) 
\big]  \;,
\end{equation}
which, introducing the Fourier transform of ${\mathcal  H}'_I$ can be
conveniently expressed as follows: 
\begin{align}
A_2 &=\; \frac{i}{2} \, 
\int_{-\infty}^{+\infty} \frac{d\nu}{2 \pi} \, \frac{1}{\nu \,-\, i 0^+} \, [ \widetilde{{\mathcal  H}'}_I(\nu),
\widetilde{{\mathcal  H}'}_I(-\nu) ] \nonumber\\
&=\; i\, 
\int_{-\infty}^{+\infty} \frac{d\nu}{2 \pi} \; \frac{1}{\nu} \; \widetilde{{\mathcal  H}'}_I(\nu) \,
\widetilde{{\mathcal  H}'}_I(-\nu)  \;,
\end{align}
where
\begin{equation}
\widetilde{{\mathcal  H}'}_I(\nu) \;=\;
\left( 
\begin{array}{cc}
\tilde{\eta}(\nu)  & \tilde{\eta}(\nu -2 \omega_0)  \\
- \tilde{\eta}(\nu + 2 \omega_0 ) & - \tilde{\eta}(\nu) 
\end{array}
\right) \;.
\end{equation}
We see that the matrix elements of $A_2$ are then given by:
\begin{align}
	& [A_2]_{11} \;=\; i \int_{-\infty}^{+\infty} \frac{d\nu}{2\pi} \,
|\tilde{\eta}(\nu)|^2 \, \frac{ 2 \omega_0}{ \nu^2 - (2\omega_0)^2 }
\nonumber\\
	& [A_2]_{22} \;=\; -[A_2]_{11} \nonumber\\
	& [A_2]_{12} \;=\;  i \int_{-\infty}^{+\infty} \frac{d\nu}{2 \pi} \;
\tilde{\eta}(\nu - \omega_0) \, \tilde{\eta}(-\nu - \omega_0 ) \, \frac{ 2
\omega_0}{ \nu^2 - \omega_0^2 } \nonumber\\
	& [A_2]_{21} \;=\; - i \int_{-\infty}^{+\infty} d\nu \;
\tilde{\eta}(\nu +  \omega_0) \tilde{\eta}(-\nu + \omega_0 ) \, \frac{ 2
\omega_0}{ \nu^2 - \omega_0^2 } \;.
\end{align}

Before proceeding to study the general case, let us consider the special
case of a harmonic $\eta(t)$, namely:
\begin{equation}
\eta(t) \,=\, \xi \, \cos(\omega t) \;, \;\;
\xi = \frac{\epsilon_0}{2 \omega_0} \,
\end{equation}
\begin{equation}
\tilde{\eta}(\nu) \,=\, \pi \xi \, [ \delta(\nu - \omega) +
\delta(\nu + \omega) ] \;.
\end{equation}
In this case, we see that:
\begin{align}
	& [A_2]_{12} \;=\; - i \pi \frac{\xi^2}{\omega_0} \, 
[ \delta(\omega - \omega_0) +
\delta(\omega + \omega_0) ] \;\nonumber\\
	& [A_2]_{21} \;=\; -  [A_2]_{12} \;. 
\end{align}

Thus, assuming that $\omega \simeq \omega_0$, we see that $A$ becomes anti-diagonal
up this order:
\begin{equation}
{\mathcal U}_I \;=\;e^{ \pi T \frac{\xi^2}{\omega_0}  \sigma_2}
\;,
\end{equation}
where $\sigma_2$ denotes a Pauli's matrix. Therefore there appears a new resonance, 
this time at $\omega = \omega_0$, and the imaginary part of $\Gamma$ behaves like:
\begin{equation}
{\rm Im}[\Gamma] \, \sim\, \pi T \, \frac{\xi^2}{\omega_0} \;.
\end{equation}
The exponential behavior,  now proportional to the square of the amplitude of the perturbation,
corresponds to a subleading parametric resonance, that appears when the external frequency 
equals the natural frequency of the system (the leading parametric resonance seen before is 
linear in the amplitude, and occurs when the external frequency is twice the natural frequency).


To finish this section, we show the structure of the third perturbative order. Applying the 
same technique as in the previous steps, this time to the third
term in the Magnus expansion we see, after some algebra, that:
\begin{equation}
A_3 \;=\; - \frac{i}{3} \, 
\int_{-\infty}^{+\infty} \frac{d\nu_1}{2 \pi} \, 
\int_{-\infty}^{+\infty} \frac{d\nu_2}{2 \pi} \;
\frac{1}{\nu_1} \, \Big( \frac{1}{\nu_2} -\frac{1}{\nu_1 -\nu_2} \Big)  
\;  
\widetilde{{\mathcal  H}'}_I(\nu_1)
\widetilde{{\mathcal  H}'}_I(\nu_2 - \nu_1) 
\widetilde{{\mathcal  H}'}_I(-\nu_2) \;.
\end{equation}


\subsection{A reinterpretation of the first-order term}
Let us see here how the firt-order term in the expansion can be related, and
therefore interpreted, in terms of the operatorial, interaction
picture evolution operator.  

Denoting by $\hat U_I(-T/2,T/2)$ the usual interaction-picture evolution
operator, we have
\begin{equation}
\hat U_I(T/2,-T/2)=T\exp[-i\int_{-T/2}^{T/2} dt\, \hat H'_I(t)]\, 
\end{equation}
where $T$ denotes the time-ordered product and the interaction Hamiltonian reads
\begin{equation}
 \hat H'_I(t)=\frac{1}{2}\frac{\epsilon(t)}{2\omega_0}\hat q(t)^2\, .
 \end{equation}

 To the lowest order in the Magnus expansion,
 \begin{equation}
\hat U_I(T/2,-T/2)\simeq\exp[-i\int_{-T/2}^{T/2} dt\, \hat H'_I(t)]\, .
\end{equation}
Taking now the limit $T\to (1-i\delta)\infty$, which selects the in and out
vacua for the initial and final times, respectively, and introducing the
usual creation and annihilation operators we obtain
\begin{equation}
\hat U_I(T/2,-T/2)\simeq\exp\left [-\frac{i}{2}\left((\hat
a^\dagger)^2\tilde\eta(2\omega_0)+\hat
a^2\tilde\eta(-2\omega_0)\right)\right ]\equiv S(z)\, ,
\end{equation}
which corresponds to a squeeze operator $S(z)$ with parameter $z=i\,\tilde\eta(-2\omega_0)$.

Thus, the lowest-order term in the expansion corresponds to having a
squeeze-like operator dictating the evolution, which is a Bogoliubov
transformation. 

\subsection{Comparison with the usual perturbative approach}\label{sec:pert}

Let us consider here the calculation of the effective action $\Gamma$,
focussing on its imaginary part, from the point of view of the standard
perturbative expansion in a would-be $0+1$-dimensional field theory, 
To that end, and to simplify the subsequent treatment, we work here
in the Euclidean formalism, where the (Euclidean) effective action is given
by:
\begin{equation}\label{eq:defgammae}
e^{-\Gamma} \;=\; \frac{\int \, {\mathcal  D}q \; e^{-{\mathcal
S}(q)}}{\int {\mathcal  D}q \; e^{-{\mathcal  S}_0(q)}} \;,	
\end{equation}
where we have used the same notation for Euclidean objects as for their
real time counterparts.  The full and free actions are now given,
respectively by:
\begin{align}
	{\mathcal  S}(q) &=\; \frac{1}{2} \, \int_{-\infty}^{+\infty} d\tau \, 
\left[ \dot{q}^2  \,+\, \omega^2(\tau) q^2  \right] \;,\nonumber\\
{\mathcal  S}_0(q) &=\; \frac{1}{2} \, \int_{-\infty}^{+\infty} d\tau \, 
\left[ \dot{q}^2  \,+\, \omega_0^2(\tau) q^2  \right] \;,
\end{align}
where $\tau$ denotes the imaginary time.

Thus, introducing the operators $\Delta$ and $\epsilon$ with kernels
defined by: 
\begin{align}
\Delta(\tau,\tau') &=\; \langle \tau|(-\partial_\tau^2 + \omega_0^2)^{-1}
| \tau' \rangle \,=\,
	\int_{-\infty}^{+\infty} \frac{d\nu}{2\pi} e^{i \nu (\tau - \tau')}
	\, \widetilde{\Delta}(\nu)  \;,\nonumber\\
\widetilde{\Delta}(\nu) &\equiv\, \frac{1}{\nu^2 + \omega_0^2} \;,
\end{align}
and 
\begin{equation}
\epsilon(\tau,\tau') \;=\; \langle \tau|\epsilon(\tau)| \tau' \rangle \,=\,
\epsilon(\tau) \, \delta(\tau-\tau') \;, 
\end{equation}
we see that:
\begin{equation}
	\Gamma \;=\; \frac{1}{2} {\rm Tr} \log( 1 + \Delta \epsilon) \;.
\end{equation}

Expanding $\Gamma$ in powers of $\epsilon$, we get a series $\Gamma \;=\; \Gamma^{(1)} \,+\, \Gamma^{(2)} \,+\, \Gamma^{(3)}
	\,+\, \ldots $.
By our initial assumption that $\int_{-\infty}^{+\infty} d\tau
\epsilon(\tau) = 0$, we see that the first order term $\Gamma^{(1)}$
vanishes, while the second order one, $\Gamma^{(2)}$, may be written as follows: 
\begin{align}
\Gamma^{(2)} &=\; \frac{1}{2} \int d\tau_1 \int d\tau_2 \;
\Gamma^{(2)}(\tau_1, \tau_2) \, \epsilon(\tau_1) \,
\epsilon(\tau_2) \nonumber\\
\Gamma^{(2)}(\tau_1, \tau_2) &=\, - \frac{1}{2} \, \Delta(\tau,\tau') \,
\Delta(\tau',\tau) \;.
\end{align}
The kernel $\Gamma^{(2)}(\tau_1, \tau_2)$ may be rendered as follows:
\begin{align}
	\Gamma^{(2)}(\tau_1, \tau_2) &=\,
	\int_{-\infty}^{+\infty} \frac{d\nu}{2\pi} e^{i \nu (\tau - \tau')}
\, \widetilde{\Gamma}^{(2)}(\nu) \nonumber\\
	\widetilde{\Gamma}^{(2)}(\nu) &=\, - \frac{1}{2} \,
	\int_{-\infty}^{+\infty} \frac{d\omega}{2\pi} \; 
	\widetilde{\Delta}(\omega) \, \widetilde{\Delta}(\omega+\nu) \;. 
\end{align}

The last expression, which is the $0+1$ dimensional version of a
real scalar field  one-loop diagram, may be exactly evaluated, the result
being:
\begin{equation}
	\widetilde{\Gamma}^{(2)}(\nu) \;=\; - \frac{1}{2 \omega_0 \, [
	\nu^2 + (2 \omega_0)^2 ]} \;. 
\end{equation}
Hence, the second order term in the effective action expansion becomes
\begin{equation}
\Gamma^{(2)} \;=\; - \frac{1}{4\omega_0} \,
\int_{-\infty}^{+\infty} \, \frac{d\nu}{2 \pi} \,
\frac{|\tilde{\epsilon}(\nu)|^2}{\nu^2 + (2 \omega_0)^2} \;, 
\end{equation}
which, when rotated back to real time has an imaginary part which is
determined by a single poles at $\nu =  \pm  2\omega_0$: 
\begin{equation}\label{ImGammapert}
	{\rm Im}\big[\Gamma^{(2)}\big] \;=\;  \frac{1}{4} \,
	\Big( \frac{|\tilde{\epsilon}(2 \omega_0)|}{2 \omega_0} \Big)^2 \;. 
\end{equation}

Of course, the last equation may be interpreted as reflecting the property
that the imaginary part appears when the frequency of $\tilde{\epsilon}$ is
sufficient to put on shell two free harmonic oscillator modes.  Note that Eq.\eqref{ImGammapert}
coincides with the lowest order Magnus approximation Eq.\eqref{resImGamma}
when this is expanded up to quadratic order in $\epsilon(t)$.

\section{The CTP effective action}\label{sec:ctp}
In situations where the main interest is to analyze the dynamical evolution
of the system (see various examples in Ref. \cite{examplesCTP}), it is relevant to compute the CTP or {\em in-in\/}
effective action, defined as \cite{ctp} 
\begin{equation}\label{CTPdef}
e^{i\Gamma_{\rm CTP}[\omega_+,\omega_-]}=\sum_n\langle0_{\rm in}\vert n\rangle_{\omega_+}\langle n \vert 0_{\rm in}\rangle_{\omega_-}.
\end{equation}
Note that the matrix elements are evaluated on two different evolutions
$\omega_\pm(t)$ of the time dependent frequency.
In a cosmological context, the frequencies are in turn functions of the scale factor of the universe $a_\pm$. The variation
of the effective action with respect to $a_+$, evaluated on $a_+=a_-=a$,
gives the semiclassical Einstein equation that takes into account the
backreaction of the quantum field on the scale factor $a$.  
One can show that $\Gamma_{\rm CTP}$ can be written in terms of the Bogoliubov
coefficients $(\alpha_\pm,\beta_\pm)$ associated to
both evolutions as follows \cite{CalHu94}
\begin{equation} \label{CTPBog}
\Gamma_{\rm CTP}[\omega_+,\omega_-]=\frac{i}{2}\log[\alpha_-\alpha^*_+-\beta_-\beta_+^*]
\end{equation}

The contribution of the CTP effective action to the equation of motion of the external degree of freedom is
\begin{equation}
\frac{\delta\Gamma_{\rm CTP}}{\delta\eta(t)}\vert_{\eta_+=\eta_-}
=\frac{i}{2}(\alpha\frac{\delta\alpha^*}{\delta\eta(t)}-
\beta\frac{\delta\beta^*}{\delta\eta(t)})\, .
\end{equation}

By its very definition, the CTP effective action produces real and causal equations of motion. It is instructive to recall that, as easily seen from the equation above,  reality comes from the identity
\begin{equation}\label{unit}
\vert\alpha\vert^2-\vert\beta\vert^2=1\,.
\end{equation}
This is in turn equivalent to the unitarity of the evolution operator $\mathcal U$. Causality, on the other hand, is a consequence of the usual composition law of the evolution operator 
\begin{equation}\label{complaw}
\mathcal U(t_f,t_i)=\mathcal U(t_f,t)\mathcal U(t,t_i)\, .
\end{equation}
We will prove this below, for an alternative derivation
see \cite{CalHu94}.

It is in general not possible to evaluate the CTP effective action and its contribution to the equations of motion for arbitrary
$\omega_\pm$.  Previous calculations are based on perturbative
approximations,
or evaluate the effective action in particular backgrounds. 
Under parametric resonance, one can work within the multiple scale analysis, although to use this approach it is necessary to assume a particular form for the external frequency. 
As we will see, when handled with care, the Magnus approximation becomes a useful tool to provide 
an analytic expression for the effective action and the
associated equation of motion. 

Let us see how the Magnus expansion above leads to the Bogoliubov
transformation that connects the {\em in\/} and {\em out\/} basis, for the
same physical system. Indeed, introducing the Bogoliubov transformation connecting the
{\em in\/} and {\em out\/} basis,
\begin{equation}
	\left(
		\begin{array}{c}
			\hat{a}_{\rm out} \\
                        \hat{a}^\dagger_{\rm out}
		\end{array}
	\right) 	
	\;=\; U \left(
		\begin{array}{c}
			\hat{a}_{\rm in} \\
                        \hat{a}^\dagger_{\rm in}
		\end{array} \right) 
\end{equation}
with 
\begin{equation}
U \;=\; 
\left(
		\begin{array}{cc}
			\alpha & \beta^*  \\
			\beta  & \alpha^*
		\end{array}
\right) \;,
\end{equation} 
since the quantum evolution for the $\hat{a}$ and
$\hat{a}^\dagger$ is identical to the one of their classical counterparts,
we see, from the first-order Magnus calculation, that:
\begin{equation}
	\alpha \,=\, {\rm cosh}[|\tilde{\eta}(2 \omega_0)|] \;,\;\;
	\beta \,=\, i \, e^{i \phi} \, {\rm sinh}[|\tilde{\eta}(2
	\omega_0)|] \;.
\end{equation}
with $\phi = {\rm arg}[\tilde{\eta}(2 \omega_0)]$. We see that the
coefficients satisfy the proper relation
Eq.\eqref{unit} to be a Bogoliubov
transformation. Besides, using Eq.\eqref{resImGamma} we see that the relation between the {\it in-out} effective action and
the Bogoliubov coefficient $\alpha$ is also satisfied:
\begin{equation}
	e^{-2 {\rm Im}\Gamma}=\vert \langle 0_{\rm out}\vert
0_{\rm in}\rangle\vert^2=\frac{1}{\vert \alpha\vert} \;.
\end{equation}

In order to compute the CTP effective action, the Bogoliubov coefficients should be computed 
for two different evolutions $\omega_\pm(t)$. Relaxing the condition of vanishing  temporal average
of $\eta_\pm(t)$ we obtain 
\begin{eqnarray}
\alpha_\pm &=&\cosh\tilde{\eta}_\pm-i\tilde\eta_\pm(0)
\frac{\sinh\tilde{\eta}_\pm}{\tilde{\eta}_\pm}\nonumber\\
\beta_\pm &=& i\tilde\eta_\pm(2\omega_0)
\frac{\sinh\tilde{\eta}_\pm}{\tilde{\eta}_\pm}\, ,
\end{eqnarray}
where we introduced the notation
\begin{equation}\label{BogMag}
\tilde\eta_\pm=\sqrt{\vert\tilde\eta_\pm(2\omega_0)\vert^2-\vert\tilde\eta_\pm(0)\vert^2 }\, .
\end{equation}
Note that in previous sections we assumed $\tilde\eta_\pm(0)=0$.
Inserting Eq.\eqref{BogMag} into Eq.\eqref{CTPBog} one easily finds an analytic expression for the CTP effective action. Interestingly enough, $\Gamma_{\rm CTP}$ depends functionally on $\eta_\pm$ through the Fourier transforms $\tilde\eta_\pm(2\omega_0)$ and  $\tilde\eta_\pm(0)$. 
From Eq. \eqref{CTPBog}  it is direct to write the real and imaginary parts of $\Gamma_{\rm CTP}$, which are related with dissipation and noise (fluctuations), respectively. In the present example we have 

\begin{equation}
{\mbox Re} \Gamma_{\rm CTP} = -\frac{1}{2} \frac{{\tilde \eta}_+(0) {\tilde \eta}_- \cosh{\tilde \eta}_- \sinh{\tilde \eta}_+ - {\tilde \eta}_-(0) {\tilde \eta}_+ \cosh{\tilde \eta}_+ \sinh{\tilde \eta}_-}{ {\tilde \eta}_+ {\tilde \eta}_- \cosh{\tilde \eta}_- \cosh{\tilde \eta}_+ - 
\sinh{\tilde \eta}_+ \sinh{\tilde \eta}_- \left( {\tilde \eta}_-(2\omega_0) {\tilde \eta}_+(2\omega_0) - {\tilde \eta}_-(0) {\tilde \eta}_+(0)\right)}, 
\end{equation}

\begin{eqnarray}
{\mbox Im} \Gamma_{\rm CTP} &=& \frac{1}{2}  \left\{\left[ {\tilde \eta}_+ {\tilde \eta}_- \cosh{\tilde \eta}_- \cosh{\tilde \eta}_+ - 
\sinh{\tilde \eta}_+ \sinh{\tilde \eta}_- \left( {\tilde \eta}_-(2\omega_0) {\tilde \eta}_+(2\omega_0) - {\tilde \eta}_-(0) {\tilde \eta}_+(0)\right)\right]^2 \right.\nonumber \\
&&+ \left. \left[{\tilde \eta}_+(0) {\tilde \eta}_- \cosh{\tilde \eta}_- \sinh{\tilde \eta}_+ - {\tilde \eta}_-(0) {\tilde \eta}_+ \cosh{\tilde \eta}_+ \sinh{\tilde \eta}_- \right]^2 \right\}^{\frac{1}{2}}   \frac{1}{{\tilde \eta}_+{\tilde \eta}_-}  .
\end{eqnarray}

The field equations can also be obtained from $\Gamma_{\rm CTP}$. However, the result is real but non-causal. These facts are a consequence of the properties of the Magnus approximation: while it respects unitarity of the evolution operator, it does not satisfy the composition law Eq.\eqref{complaw}. 

The problem of the non causality of the equation of motion can be solved by applying the Magnus approximation after taking the variation of the action with respect to $\eta_\pm$. 
Let us write
\begin{equation}\label{complaw2}
\mathcal U(T/2,-T/2)=\mathcal U(T/2,t')\mathcal U(t',-T/2)
\equiv \mathcal U_2\mathcal U_1 \, ,
\end{equation}
with $-T/2<t<t'<T/2$ and $t$ is the time at which we want to
evaluate the equation of motion. The equation of motion can be
written as
\begin{equation}
\frac{\delta\Gamma_{\rm CTP}}{\delta\eta(t)}\vert_{\eta_+=\eta_-}
= -\mathcal U_1^{-1}\frac{\delta U_1}{\delta\eta(t)}\vert_{11}
\, .
\end{equation}
Taking the limit $t'\to t$, we see that the equation of motion depends only  the values of $\eta(\tau)$ with $\tau \leq t$. 
This proves that the validity of the composition law implies causality. Moreover, we can now use the Magnus approximation
to evaluate $\mathcal U_1$, and in this way we assure causality.

The effective action $\Gamma_{\rm CTP}$ depends, in the Magnus approximation, on the Fourier transform of $\eta(t)$. When the Magnus approximation is applied to the equation of motion,
it depends on the ``retarded''  Fourier transform
\begin{equation}
{\tilde\eta}_{\rm ret}(\omega)=\int_{-\infty}^t d\tau\, \eta(\tau)e^{-i\omega \tau}\, ,
\end{equation}
evaluated at $\omega=2\omega_0$ and $\omega=0$. Explicitly,
we have that the equation of motion can be written as
\begin{eqnarray}
\frac{\delta\Gamma_{\rm CTP}}{\delta\eta(t)}\vert_{\eta_+=\eta_-} &=&
-\frac{1}{2} \frac{\sinh{\tilde \eta}_{\rm ret} \cosh{\tilde \eta}_{\rm ret} }{{\tilde \eta}_{\rm ret} } - \frac{1}{2} \frac{\sinh{\tilde \eta}_{\rm ret}}{{\tilde \eta}_{\rm ret}} {\mbox Im}\left[ {\tilde \eta}_{\rm ret}(2\omega_0) e^{-2 i \omega_0 t}\right] \\
&-& \frac{1}{2} \frac{{\tilde \eta}_{\rm ret}(0)}{{\tilde \eta}_{\rm ret}^2} \left( 1 - \frac{\sinh{\tilde \eta}_{\rm ret} \cosh{\tilde \eta}_{\rm ret} }{{\tilde \eta}_{\rm ret}} \right) \left( {\mbox Re} 
\left[ {\tilde \eta}_{\rm ret}(2\omega_0) e^{-2 i \omega_0 t} \right] - {\tilde \eta}_{\rm ret}(0) \right), \nonumber 
\end{eqnarray} with $\tilde\eta_{\rm ret}=\sqrt{\vert\tilde\eta_{\rm ret} (2\omega_0 )\vert^2-\vert\tilde\eta_{\rm ret} (0)\vert^2 }$.
Taking into account that $\tilde\eta_{\rm ret}$ is either a real or a  pure imaginary number, the reality and causality of the equation of motion 
is evident. 

\section{Conclusions}\label{sec:conc}
We have calculated the in-out effective action for a single harmonic
oscillator with a time-dependent frequency, applying the Magnus expansion
to the solution of the homogeneous second-order equation which yields the
result for the fluctuation determinant.  We evaluated explicitly the first
and second order terms in that expansion, and presented the structure of
the third order one. By studying the imaginary part of the corresponding
terms in the effective action, we have shown that it automatically captures
non-trivial features, like the position of the (parametric) resonances,
which is a crucial aspect, for example, to study particle creation in
dynamical systems (for example, in nonstationary cavity quantum
electrodynamics, such as those cases related with the dynamical Casimir
effect). The same calculation allowed us to compute the unitary matrix that
implements the Bogoliubov transformation between the {\em in} and {\em out}
basis. Up to each order in the expansion, the transformation is unitary.

We have compared the results of the Magnus approach to the in-out effective action with
the ones one would obtain by applying standard, field-theoretic perturbation
theory in Euclidean time, showing that they agree to the lowest order.  
The Magnus expansion, however, provides a non-trivial resummation of the
perturbative results,  which preserves the unitary evolution. For example,
to the lowest order, that expansion amounts to including the evolution
dictated by a squeeze operator, with a parameter determined by the Fourier
transform on the frequency at its first resonance.

We have also considered the CTP effective action, implementing the
Magnus approximation at the level of the equations of motion.  This
effective action, which is written in terms of the Bogoliubov coefficients,
is relevant in order to study dynamical evolutions as non-equilibrium
problems, open quantum systems, etc.  We have shown the Magnus
approximation becomes a useful tool to provide an analytic expression for
the effective action and the associated equation of motion. To lowest order
in the Magnus expansion, the equation of motion involves the ``retarded"
Fourier transform of the perturbation, evaluated at twice the natural
frequency of the oscillator. This is a simple way to take
into account the back reaction of the parametric resonance on the external pumping.

\section*{Acknowledgements}
This work was supported by ANPCyT, CONICET, UBA and UNCuyo; Argentina. 


\begin{thebibliography}{bib}
 \bibitem{Birrell} N. D. Birrell and P. C. W. Davies, {\it Quantum Fields in
Curved Space} (Cambridge University Press, Cambridge, 1982).
\bibitem{Reviews}
  V.~V.~Dodonov,
  Phys.\ Scripta {\bf 82} (2010) 038105;
  D.~A.~R.~Dalvit, P.~A.~Maia Neto and F.~D.~Mazzitelli,
  Lect.\ Notes Phys.\  {\bf 834} (2011) 419;
  P.~D.~Nation, J.~R.~Johansson, M.~P.~Blencowe and F.~Nori,
  Rev.\ Mod.\ Phys.\  {\bf 84} (2012) 1.
  
  \bibitem{nos1}  F. C. Lombardo and F. D. Mazzitelli, Physica Scripta {\bf 82}, 038113 (2010).
 
  
  \bibitem{nos2} M. Crocce, D.A.R. Dalvit, and F.D. Mazzitelli,  Phys. Rev. A {\bf 64}, 013808 (2001); D.A.R. Dalvit and  F.D. Mazzitelli, Phys. Rev. A {\bf 59}, 3049 (1999); M. Crocce, D.A.R. Dalvit, F.C. Lombardo, and F.D. Mazzitelli, Phys. Rev. A {\bf 70}, 033811 (2004); P.I. Villar, A. Soba, and F.C. Lombardo
Phys. Rev. A {\bf 95}, 032115 (2017). 
  
\bibitem{CalHu94} E.~Calzetta and B.~L.~Hu,
  Phys.\ Rev.\ D {\bf 49}, 6636 (1994).

\bibitem{nos3} F.C. Lombardo and  F.D. Mazzitelli, Phys. Rev. D {\bf 55}, 3889 (1997).

 \bibitem{Magnus}
  W. Magnus, Comm. Pure and Appl. Math. {\bf VII}, 649 (1954); for a
review see S. Blanes, F. Casas, J.A. Oteo and J. Ros, Phys. Rep. {\bf 470},
151 (2009).

\bibitem{11} W.R. Salsman, J. Chem. Phys. {\bf 82}, 822 (1985).

\bibitem{12} Luca D'Alessio and Anatoli Polkovnikov, Annals of Physics {\bf 333}, 19 (2013).

\bibitem{viola} Lorenza Viola, Emmanuel Knill and Seth Lloyd, Phys. Rev. Lett. {\bf 82}, 2417 (1999).

\bibitem{parametric} B. Zhu, T. Rexin, and L. Mathey, Zeitschrift f\"ur Naturforschung A {\bf 71}, 921 (2016).

\bibitem{bloch} Tobias Lipfert, Dmitri Horoshko, Giusepe Patera, and Mikhail Kolobov, arXiv:1806.09384[quant-ph].

\bibitem{sipe} N. Quesada and J.E. Sipe. Phys. Rev. Lett. {\bf 114}, 093903, (2015).

\bibitem{2LS} P. Nalbach and V. Leyton, arXiv:1806.02738[quant-ph].

\bibitem{thimmel} B. Thimmel, P. Nalbach, and O. Terzidis, Eur. Phys. J. B {\bf 9}, 207 (1999).

 \bibitem{Gelfand} 
I.~M.~Gelfand and A.~M.~Yaglom,
J.\ Math.\ Phys.\  {\bf 1}, 48 (1960).

\bibitem{pedagogico} S. Blanes, F. Casas, J.A. Oteo and J. Ros, Eur. J. Phys. {\bf 31}, 907 (2010). 

  \bibitem{Bender} C.M. Bender and S.A. Orszag, {\it Advanced Mathematical Methods for Scientists and Engineers}
  (Mc Graw-Hill, Inc, New York, 1978), Chapter 11.
  
  \bibitem{examplesCTP} F. Lombardo, F.D. Mazzitelli, 
Phys. Rev. D {\bf 53} (2001); F.C. Lombardo abd  D.L\'opez Nacir
Phys. Rev. D {\bf 72}, 063506 (2005); F.C. Lombardo and P.I. Villar
Phys. Letts. A {\bf 336} 16 (2005); N.D. Antunes, F.C. Lombardo, D. Monteoliva, and P.I. Villar
Phys. Rev. E {\bf 73}, 066105 (2006); C.D. Fosco, F.C. Lombardo, and F.D. Mazzitelli
Phys. Rev. D {\bf 76}, 085007 (2007);  A.E.Rubio L\'opez 	and  F.C. Lombardo
Phys. Rev. D {\bf 89}, 105026 (2014); M.B. Far\'\i as and  F.C. Lombardo
Phys. Rev. D {\bf 93}, 065035 (2016).
  
  \bibitem{ctp} J. Schwinger, J. Math. Phys. (N.Y.) {\bf 2}, 407 (1961); L. V. Keldysh, Zh. Eksp. Teor. Fiz. {\bf 47}, 1515 (1965) [Sov. Phys. JETP {\bf 20},	1018	(1965)]; E. A. Calzetta	and	B. L.	Hu, Nonequilibrium Quantum Field Theory (Cambridge University Press, Cambridge, England, 2008).

  
\end{thebibliography}
\end{document}